# Science Citation Index data: Two additional reasons against its use for administrative purposes

J. Marvin Herndon
Transdyne Corporation
San Diego, CA 92131 USA

mherndon@san.rr.com

http://UnderstandEarth.com

February 14, 2007

**Abstract:** First, for decades, the use of anonymity in reviews for science funding proposals and for evaluating manuscripts for publication has been gradually corrupting American science, encouraging and rewarding the dark elements of human nature. Unethical reviewers, secure and unaccountable through anonymity, all too often make untrue and/or pejorative statements to eliminate their professional competitors. Survival in this corrupt environment has led to a consensus-only mentality. Consequently, important scientific contradictions, if they can be published at all, are selectively ignored in many instances out of fear of anonymous retaliation. Science Citation Index data in such a corrupt environment may be of little administrative value, except for possible use in documenting scientific fraud. Second, as knowledge of the administrative use of Science Citation Index data spreads, scientists will adapt and will shift to research on popular subjects to elicit greater numbers of citations, rather than to take the paths less trodden where important scientific discoveries may lay waiting.

The fascinating and insightful debate concerning the use and misuse of Science Citation Index (SCI) data as an administrative tool [1-6] continues with the most recent contribution from Rustum Roy [7] warning of considerable dangers associated with the frequently improper use of such data. Roy articulates well many of the major problems associated with the use of SCI data of which I am in full agreement. I would like to mention two, however, that he does not address directly, one pertaining to the questionability of what is actually being measured by the data, the other bearing on the deleterious effect its use may have on the progress of science.

In 1951, the U. S. National Science Foundation was established to provide support for post-World War II scientific research. Soon thereafter, someone had the idea that reviewers of scientific proposals for government grant monies should be anonymous; the idea being that anonymity would encourage honesty in evaluation even when those reviewers might be competitors or might have vested interests. The idea of using anonymous reviewers was also rapidly adopted by many editors of scientific journals. (Prior to World War II, when a scientist wanted to publish a paper, he/she would send it to the editor of a scholarly journal for publication and generally it would be published. A new, unpublished scientist was required to obtain the endorsement of a published scientist before submitting a manuscript.)

There is a major flaw in the blanket application of anonymity. If anonymity leads to greater truthfulness, then it could be put to great advantage in the courts. Courts have in fact utilized anonymity — in the infamous Spanish Inquisition and in virtually every totalitarian regime — and the results are always the same: People denounce others for a variety of reasons and corruption becomes rampant.

For decades, the use of anonymity within the National Science Foundation, NASA, and elsewhere has been gradually corrupting American science. Unethical reviewers — secure, camouflaged, masked and hidden through anonymity — all too often make untrue and/or pejorative statements to eliminate their professional competitors. It is a pervasive, corrupt system that encourages and rewards the darker elements of human nature. Under adverse conditions, humans adapt to their environment if they want to survive. And, survival in this corrupt environment has led to a "consensus only" mentality. Scientists are quick to realize that citing work that challenges the "consensus view" might well result in their own reports not being published and their proposals for grant aid receiving only lukewarm reviews. Consequently, publications of important scientific contradictions, if they can be published at all, are selectively ignored in many instances. Science Citation Index data in such a corrupt environment may be of little administrative value, except for possible use in documenting scientific fraud.

In the 1970s, there was a movement in American universities to make use of students' evaluations of their classroom teachers and teaching assistants. In some instances, a team would come into the classroom to collect students' evaluation forms, while the teacher and teaching assistant were required to leave the room. Those evaluations would then be analyzed and used for administrative purposes, especially in promotion and tenure decisions.

People are the same worldwide. Generally, they want to earn a living and to be successful and secure in doing so. From personal experience, I know the response of some teachers to students' teacher evaluations. The teachers became less demanding, lowered their expectations, and, consequently, received more glowing reviews from many of their students. Teachers adapt and scientists adapt. As knowledge of the administrative use of Science Citation Index data spreads, scientists will adapt and will shift to research on popular subjects to elicit greater numbers of citations, rather than to take the paths less trodden where important scientific discoveries may lay waiting.

Beyond the use and misuse of Science Citation Index data, Roy [5, 7] and I [8, 9] are in agreement that emerging India should chart her own course and not simply parrot a system that has been mal-administered to the point of corruption.